 \documentclass[final,1p,times]{elsarticle}

 \usepackage{graphicx}

\usepackage{amssymb}

\journal{Nuclear Physics A}

\begin{document}

\begin{frontmatter}



\title{Hydrodynamics: Fluctuating Initial Conditions and 
  Two-particle Correlations}


\author{R.P.G. Andrade, F. Grassi, Y. Hama, and 
  W.-L. Qian\fnref{qian}\fntext[qian]
  {Now at the Instituto de Ci\^encias Exatas, Universidade
  Federal de Ouro Preto, Ouro Preto - MG, Brazil}} 

\address{Instituto de F\'{\i}sica, Universidade de S\~ao 
  Paulo, Brazil}

\begin{abstract} 
Event-by-event hydrodynamics (or hydrodynamics with fluctuating initial conditions) has been developed in 
the past few years. Here we discuss how it may help to 
understand the various structures observed in 
two-particle correlations. 

\end{abstract}


\end{frontmatter}


\section{3+1 hydrodynamics: comparison with data}

One of the most striking results in relativistic heavy ion  
collisions is the existence of  structures in the 
two-particle correlations \cite{ridge1,ridge1s,ridge2,dhump1,phobos,phoboss,ridge3}  plotted as function of the pseudorapidity difference 
$\Delta \eta$ and the angular spacing $\Delta \phi$. 
The so-called ridge has a narrow $\Delta \phi$ located around 
zero and a long $\Delta \eta$ extent. 
The other structure located opposite has a single or 
double hump in $\Delta \phi$, its $\Delta \eta$ extent is not 
well established. 

It has been suggested that the combined effect of 
longitudinal high energy density tubes (leftover from initial 
particle collisions) and transverse expansion is 
responsible for the ridge 
\cite{voloshin,shuryak,lml1,lml2,sg}. A hydrodynamics based 
explanation is attractive because of the various 
similarities between bulk matter and ridge (transverse 
momentum spectra, baryon/meson ratio, etc). 
In order to compare with the ridge data, 3+1 hydrodynamics 
must be used. Traditionally, smooth initial conditions have  
been used (see e.g. \cite{Hirano,Nonaka}).  
However some groups have started to develop an event-by-event 
approach: for each collision, some initial conditions are  
generated, hydrodynamics is run and results are stored. This  
is done many times thus mimicking experience. This method was pioneered by the Brazilian collaboration SPheRIO  \cite{spherioref1,spherioref2,spherioref3} (some typical 
results can be seen in \cite{sph1,sph2,sph3,jun}). 
It has been used also by the German group with 
H. Petersen/M. Bleicher et al.$\,$\cite{urqmdhydro}. More 
recently K. Werner and collaborators have also obtained results \cite{klaus}.

The SPheRIO group uses the NeXus code \cite{nexus} to 
generate initial conditions and the Smooth Particle 
Hydrodynamics method to solve the perfect fluid conservation 
equations. Results on two-particle correlations can be found 
in \cite{jun,ismd09,sqm09}. 
H. Petersen/M. Bleicher et al. start with initial conditions generated with UrQMD and solve the perfect fluid  
hydrodynamics equations with a standard (grid) method. 
They have no results on  two-particle correlations at RHIC. 
K. Werner et al. use EPOS initial conditions and also a 
standard (grid) method to solve the ideal hydrodynamics 
equations. Their results on two-particle correlations can be 
found in \cite{klaus}. 

\begin{figure} 
\includegraphics[width=4.cm]{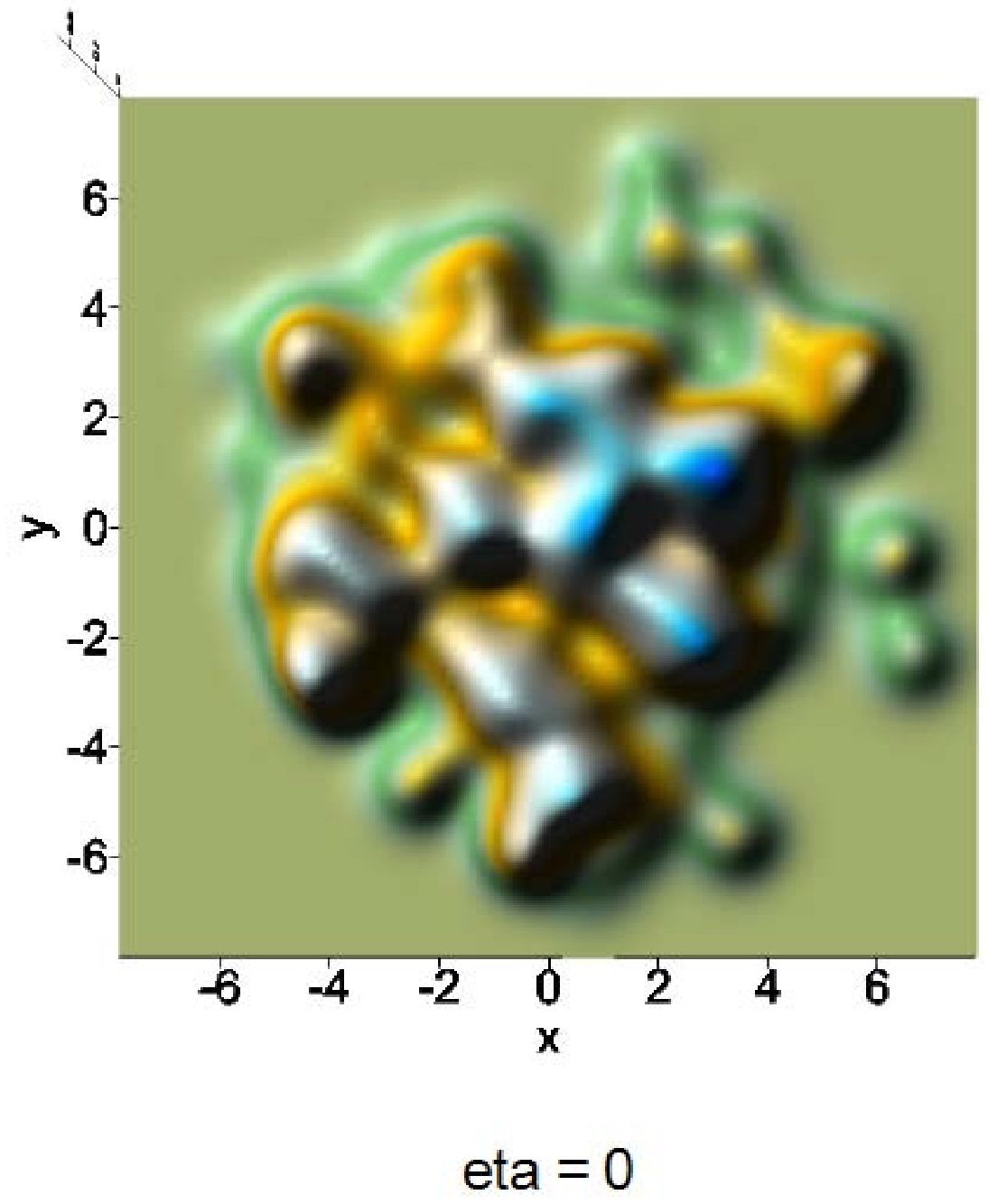}
\includegraphics[width=4.5cm]{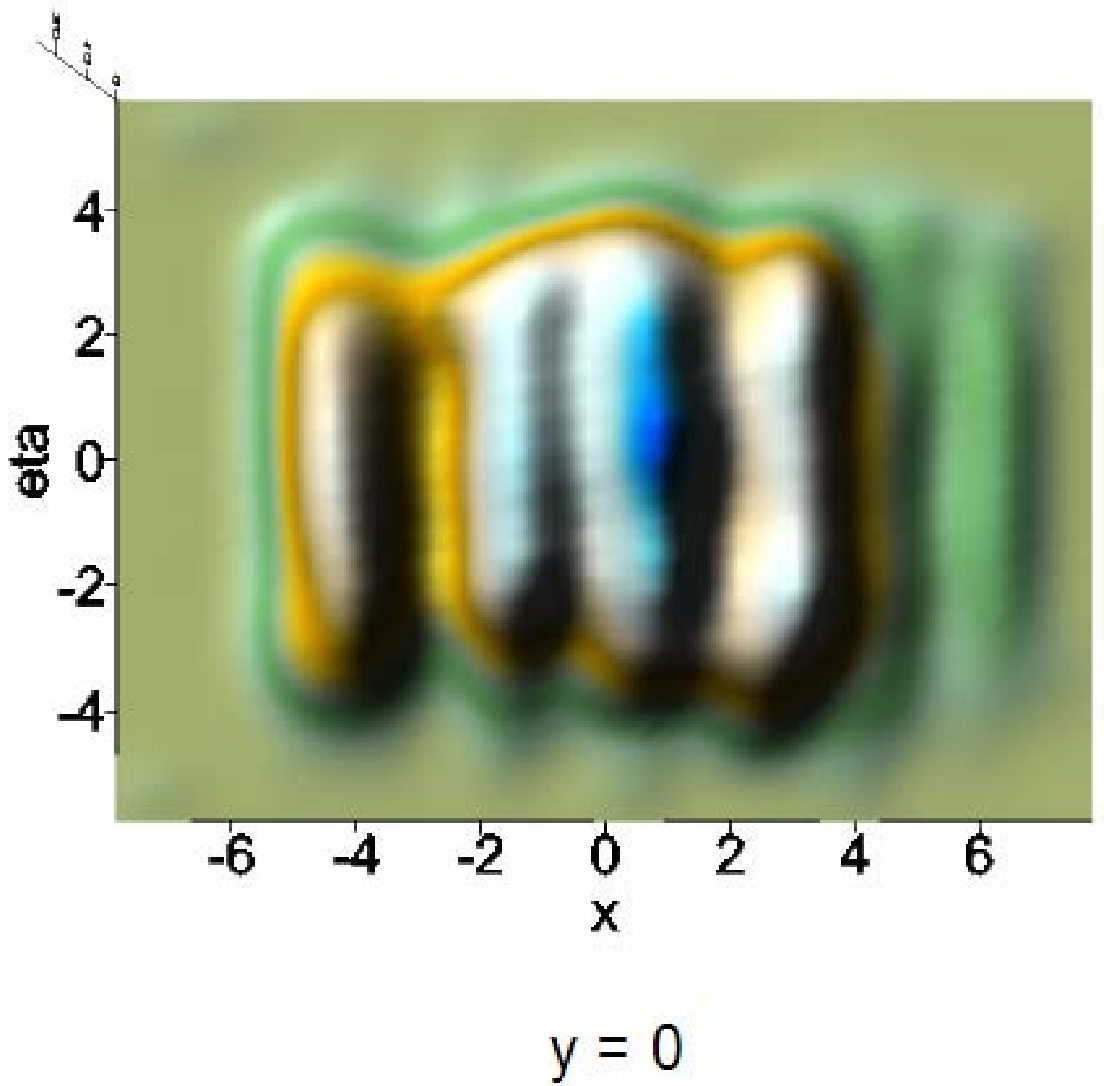}
\hspace{-.6cm} 
\includegraphics[width=5.8cm]{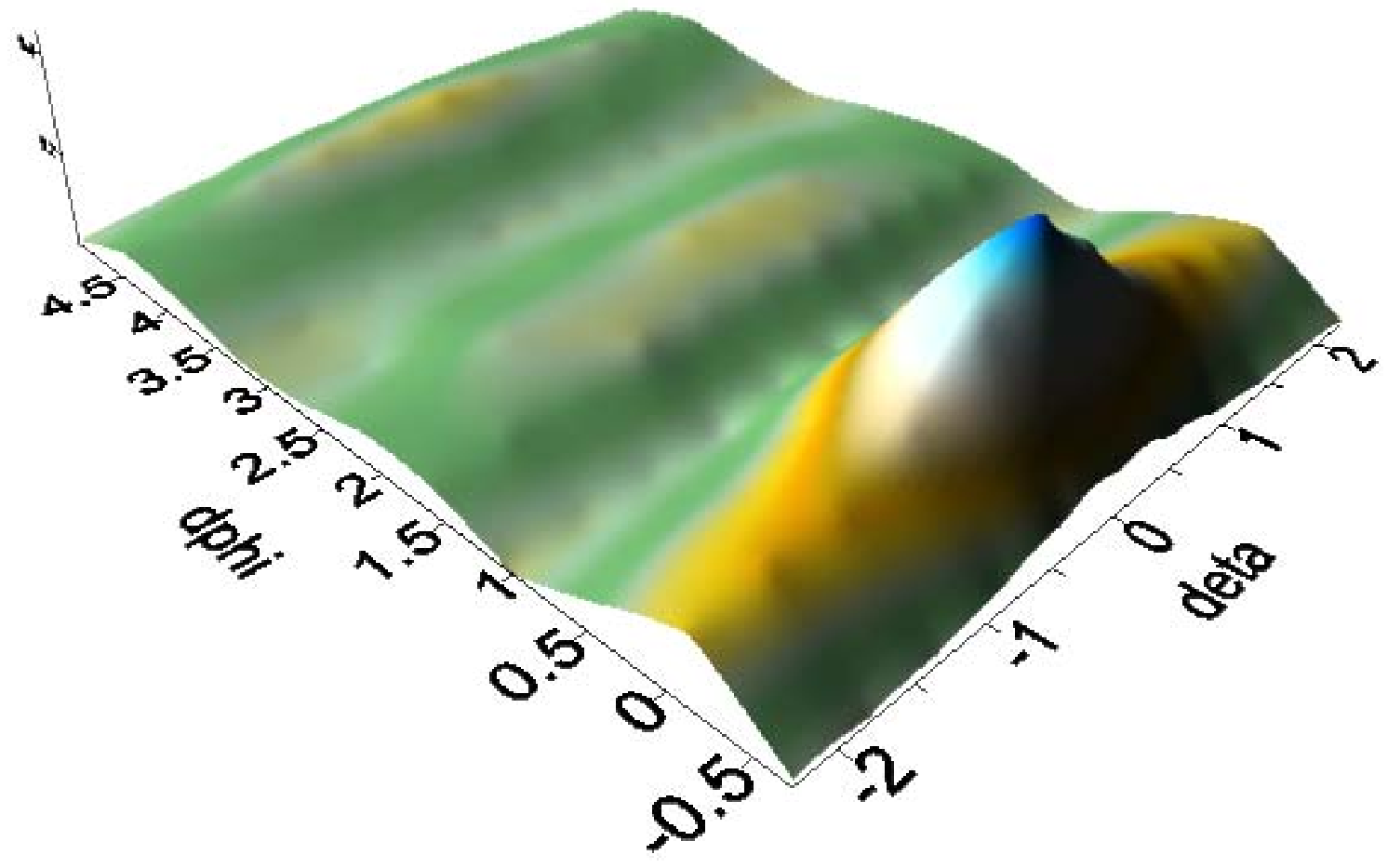}
\caption{\label{ic2d} NeXSPheRIO central Au+Au collisions at 
 200 GeV A: initial energy density and resulting two-particle
 correlation ($p_t^{trig}>2.5\,$GeV 
 $\times\ p_t^{assoc}>1.5\,$GeV).} 
\end{figure}

In the NeXSPheRIO approach, initial conditions have tubular 
structures and two particle correlations exhibit {\em near-} 
and {\em away-side} ridges as can be seen in 
Fig. \ref{ic2d}. 
As a double check, two different methods for elliptic flow  
subtraction have been used: ZYAM and event plane alignment. 
In the calculation by K. Werner et al., initial conditions 
also have tubular structures, a plane alignment method for 
elliptic flow subtraction is used  and two-particle  
correlations exhibit near-side \cite{klaus} and 
{\em small} away-side ridges \cite{KWprivate}. 

\begin{figure}[h] 
\vspace*{-.5cm} 
\begin{minipage}[h]{70mm} 
\hspace{-.4cm} 
\includegraphics*[width=7.5cm]{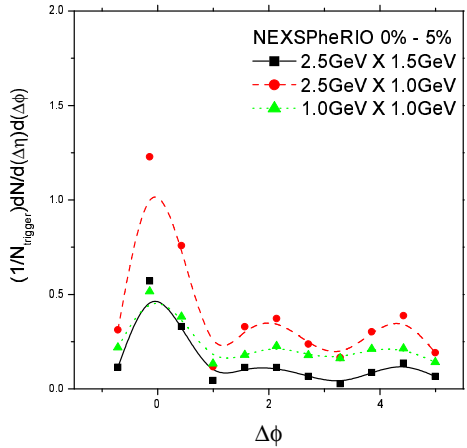} 
\vspace*{-4.5cm}
\caption{\label{pt2d} $p_t$ dependence of NeXSPheRIO 
 two-particle correlations for 0-20\% Au+Au collisions. 
 Left column in legend is trigger momentum and right, 
 associated particle momentum.} 
\end{minipage} 
\hspace{.8cm}
\begin{minipage}[t]{55mm} 
\vspace*{-4.cm} 

the single-particle angular distribution becomes 
sharper, reducing the smearing effect when integrated over 
the trigger angle. This behavior is in agreement with data 
\cite{dhump3}. 
2) When going from central to peripheral collisions, the 
near-side ridge decreases and the away-side ridge changes  
from double to single hump, as seen in figure \ref{centr2d}. 
This is in agreement with data  \cite{dhump3}. 
3) The correlation can also be studied as a function of the  
trigger-particle angle with relation to the event plane. In  
Figure \ref{inoutridge} for a mid-central window, the  
away-side ridge changes from single peak for in-plane trigger 
to double peak for out-of-plane trigger. For central 
collisions (not shown), it is always double-peaked. This is in agreement with data \cite{inout}. 
\end{minipage} 
\end{figure}

In addition to reproducing both the near and away-side 
structures, NeXSPheRIO leads to good qualitative  
agreement with various data. For example, 
1) for fixed $p_t^{trig}$ and increasing $p_t^{assoc}$, the  
near-side and away-side peaks decrease as seen in figure 
\ref{pt2d} for central collisions (this is generally expected since the number of associated particles decreases). On the 
other side, for fixed $p_t^{assoc}$ and increasing $p_t^{trig}$, the peaks increase. As shown below (see Fig. \ref{1dist}), this is because\hfilneg\ 

\begin{figure} 
\begin{center} 
\includegraphics[width=12.cm]{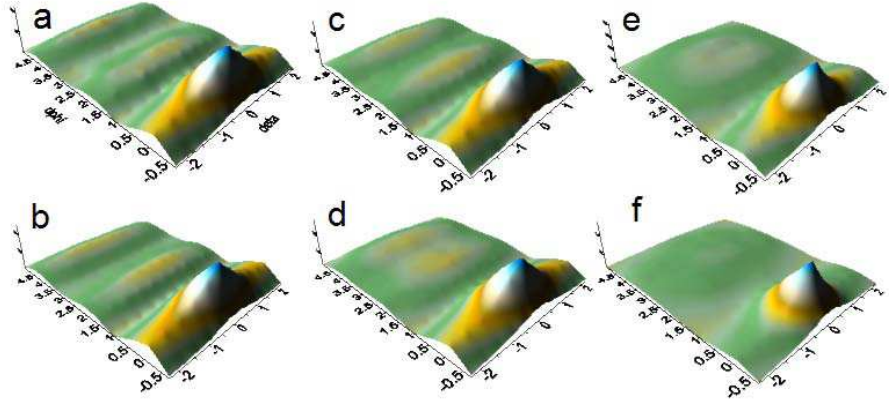}
\caption{\label{centr2d} NeXSPheRIO two-particle 
 correlations for central (a) to peripheral (f) collisions.} 
\end{center} 
\end{figure} 
\vspace*{.5cm} 

\section{2+1 hydrodynamics: one tube model}

\begin{figure}[t]
\begin{center}
\includegraphics[width=5.cm]{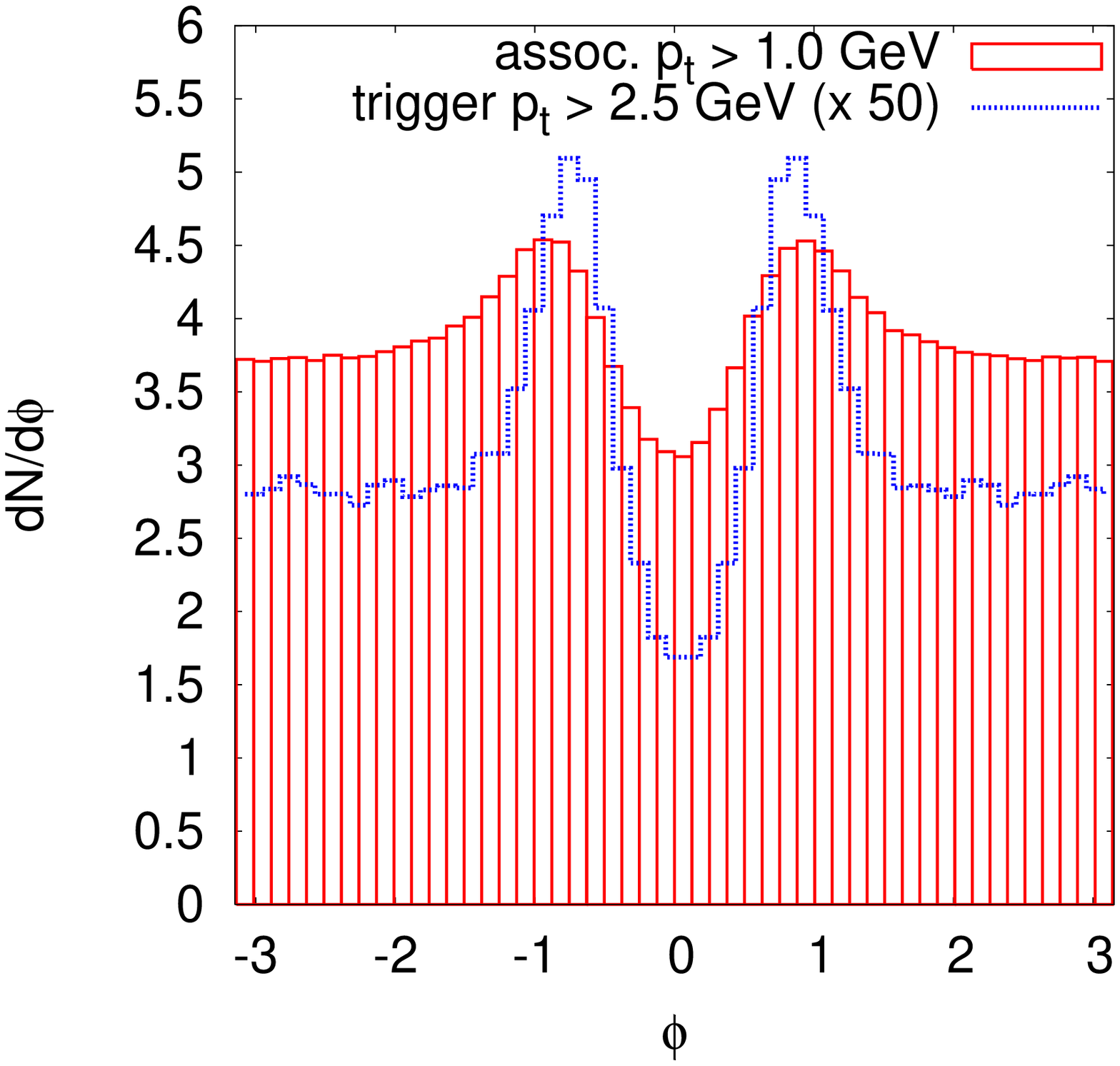}
\includegraphics[width=5.cm]{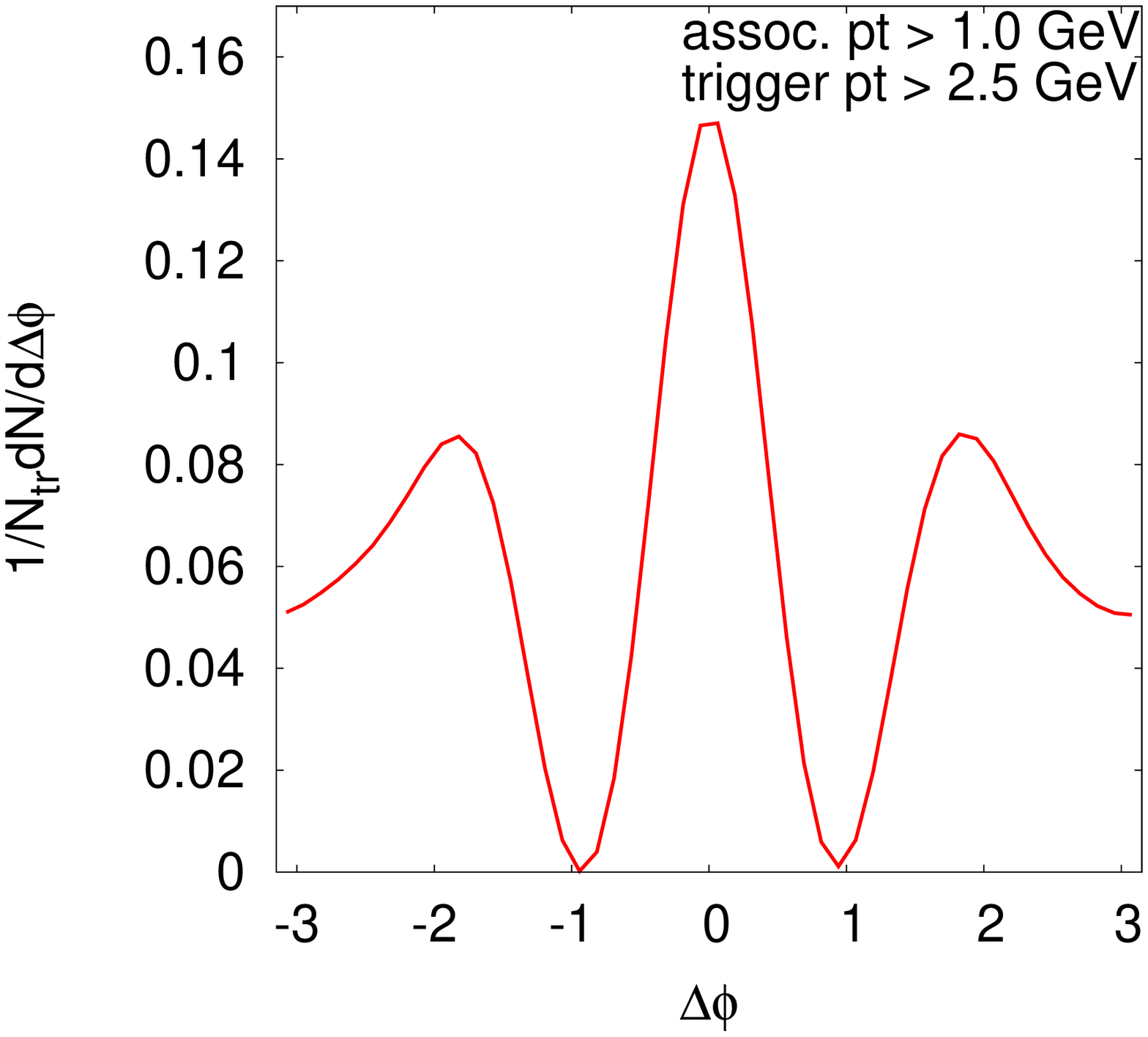}
\caption{\label{1dist}
Angular distributions of particles in some different $p_T$ intervals, in the 
simplified model and resulting two-particle correlations.} 
\end{center} 
\end{figure}

When using NeXSPheRIO, it is not clear how the various  
structures in the two particle correlations are generated. 
To investigate this, we use a simplified model. 
Fig. \ref{ic2d} (left and center) shows a typical example of 
initial conditions (initial energy density) obtained in NeXus 
with various tubular structures along the collision axis. 
\begin{figure}   
\begin{center} 
\includegraphics[width=12.cm]{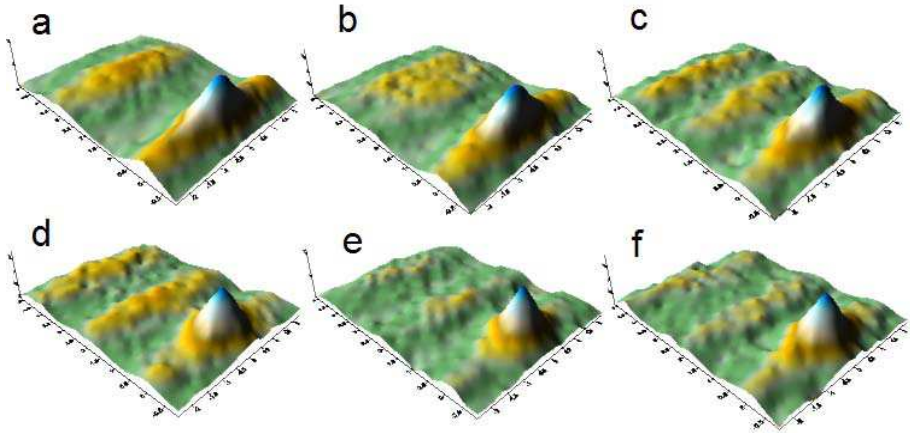}
\caption{\label{inoutridge} NeXSPheRIO (20-30\%) 
results for in-plane (a) to out-of-plane (f) trigger.} 
\end{center} 
\vspace*{-1.cm} 
\end{figure}
In the simplified model, only one of the high energy tubes  
from NeXus (chosen close to the border) is considered and 
the complex background is smoothed out by using the average 
over many events (for details see \cite{ismd09}). 
Transverse expansion is computed numerically while  
longitudinal expansion is assumed boost-invariant, until 
freeze out at some constant temperature. The resulting single 
particle angular distribution has two peaks located on both 
sides of the position of the tube placed at $\phi=0$ as seen 
in Fig. \ref{1dist}\hfilneg\  

\newpage 
\noindent (left). 
This double peak structure is observed for all transverse momenta at more or less the same position. 

\begin{figure}[b] 
 \includegraphics[width=4.6cm]{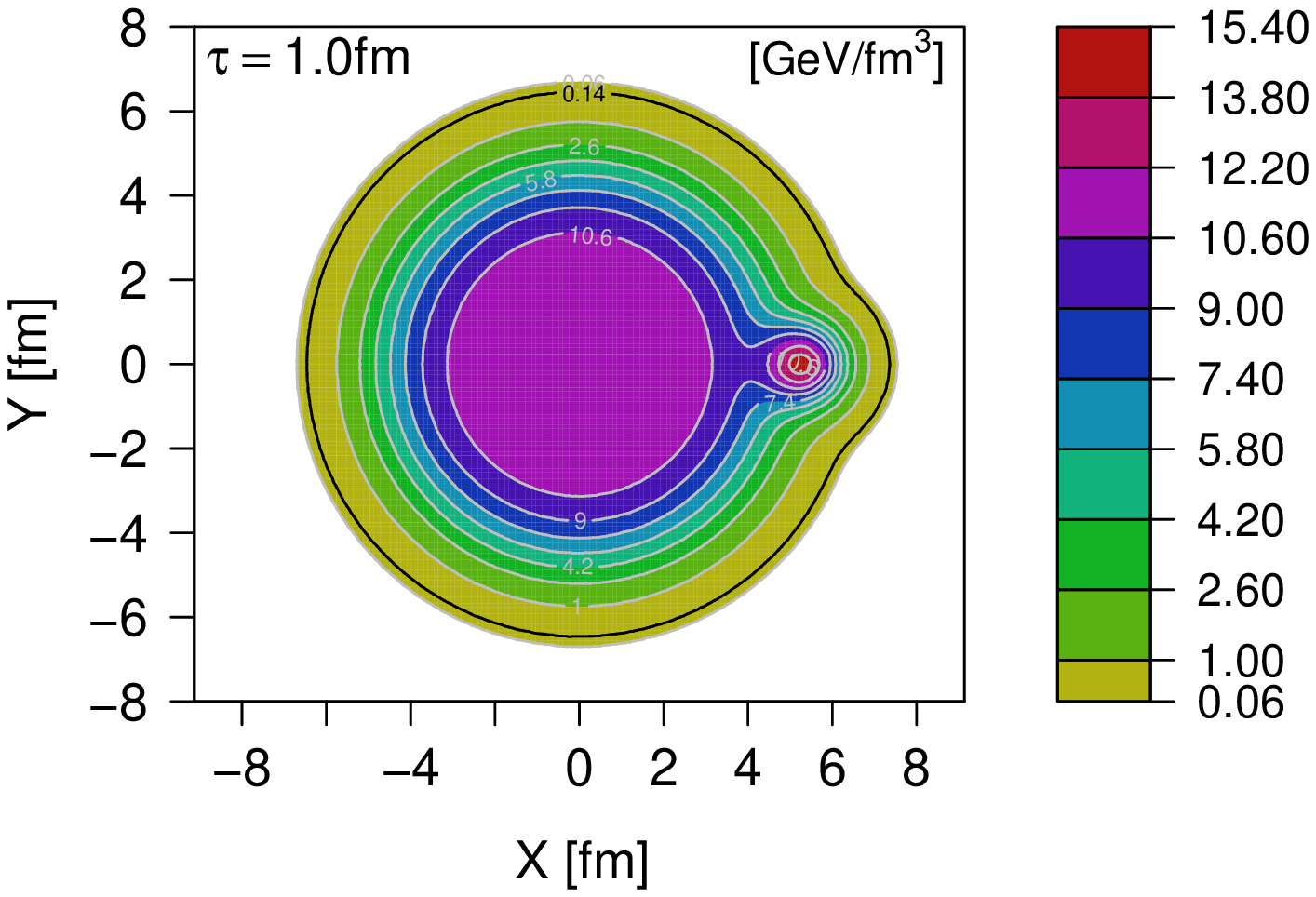}
 \hspace{-.3cm}
 \includegraphics[width=4.6cm]{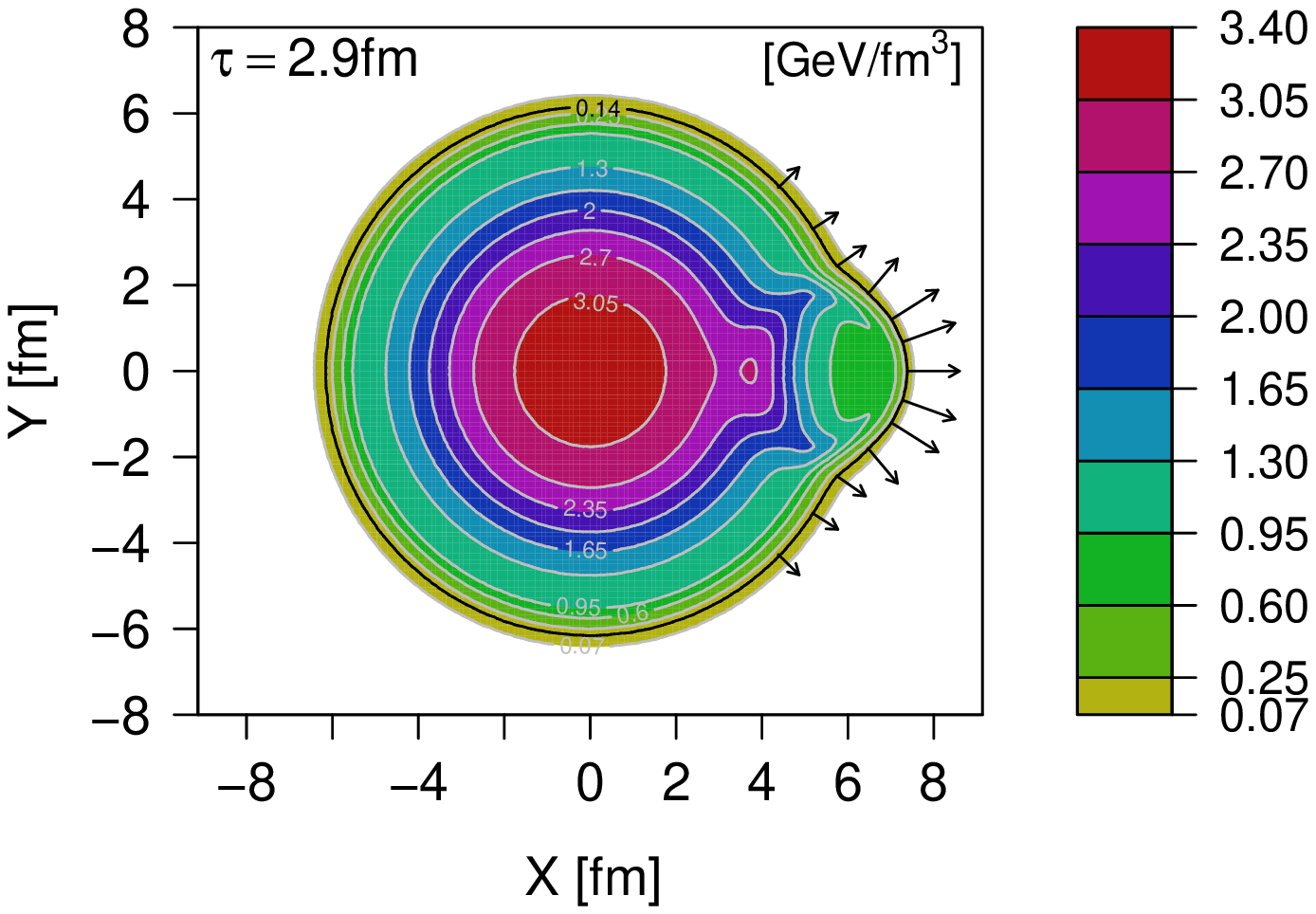}
 \hspace{-.3cm}
 \includegraphics[width=4.6cm]{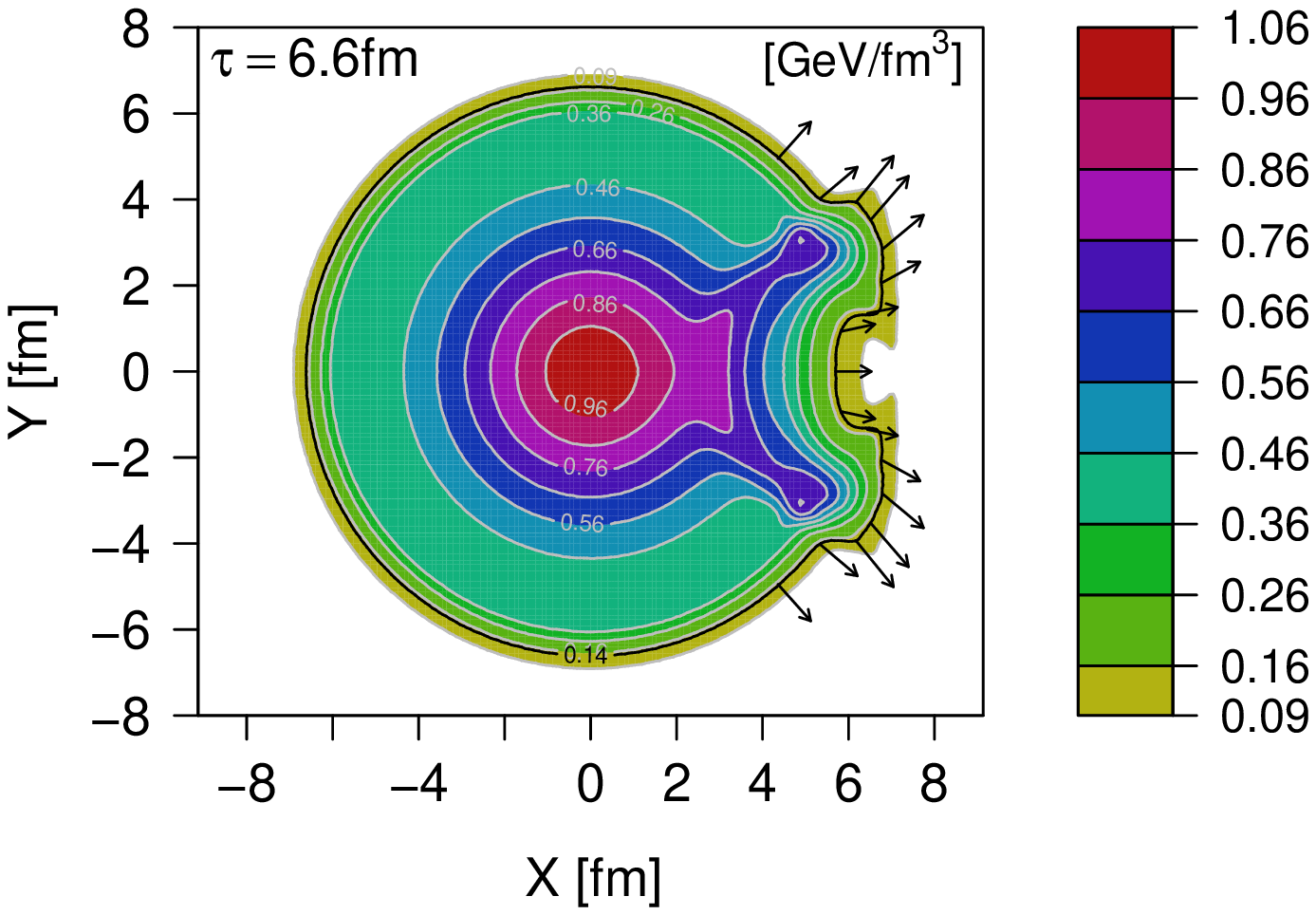}
 \caption{\label{temp} Temporal evolution of energy density 
 for the simplified model (times: 1.0, 2.9 and 6.6 fm). 
 Arrows indicate fluid velocity on the freeze out surface,  
 thicker curve labeled by the freeze out temperature 0.14
 GeV} 
\end{figure}

This two-peak emission is in contrast with what happens 
when a blast wave is assumed, namely the fact that 
high-energy tubes emit in a single direction. However its 
occurrence can be understood from Fig. \ref{temp}. As time 
goes on, as a consequence of the tube expansion, a hole 
appears at the location of the high-energy tube (as in  
\cite{shuryakhole}). This hole is surrounded by matter that  
piles up in a roughly semi-circular cliff of high energy  
density matter, guiding the flow of the background matter 
into two well-defined directions. 
The outer part of the circle does not appear 
because the vacuum does not offer any resistance against the tube expansion and this part of the fluid is emitted  
isotropically. The two extremities of the cliff emit 
more fast particles than the background, this gives rise to 
the two-peaks in the single-particle angular distribution. 
The emission is not quite radial as shown by Fig. \ref{temp}, 
indicating that there was a deflection of the background 
flow caused by the pressure put by the high-energy tube. 

From Fig. \ref{1dist} left, we can guess how the two-particle 
angular correlation will be. The trigger particle is more 
likely to be in one of the two peaks. We first choose the 
left-hand side peak. The associated particle is more likely 
to be also in this peak i.e. with $\Delta \phi=0$ or in the 
right-hand side peak with $\Delta \phi\sim +2$. If we choose 
the trigger particle  in the right-hand side peak, the 
associated particle is more likely to be also in this peak 
i.e. with $\Delta \phi=0$ or in the left-hand side peak with 
$\Delta \phi\sim -2$. So the final two particle angular 
correlation  must have a large central peak at 
$\Delta\phi=0$ and two smaller peaks respectively at 
$\Delta\phi\sim\pm2$. 
Fig. \ref{1dist} (right) shows that this is indeed the case.
The peak at $\Delta \phi=0$ corresponds to the near-side 
ridge and the peaks at 
$\Delta \phi\sim \pm 2$ form the double-hump ridge. 
We have checked that this structure is robust by studying  
the effect of the height and shape of the background, initial 
velocity, height, radius and location of the tube \cite{ismd09}. 

\begin{figure}[h] 
\hspace*{-.2cm} 
\includegraphics[width=5.2cm]{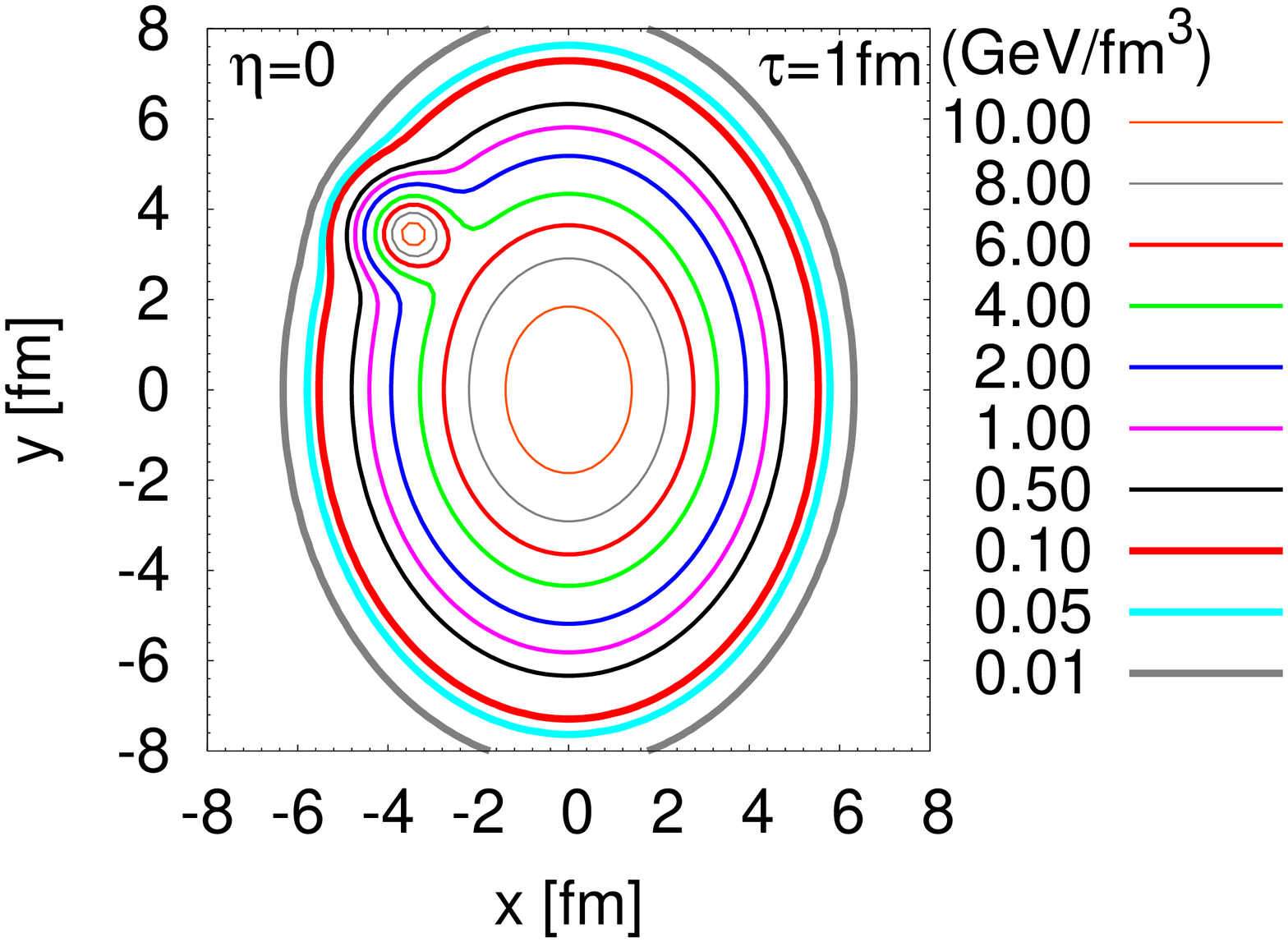}
\hspace*{-.3cm}
\includegraphics[width=4.5cm]{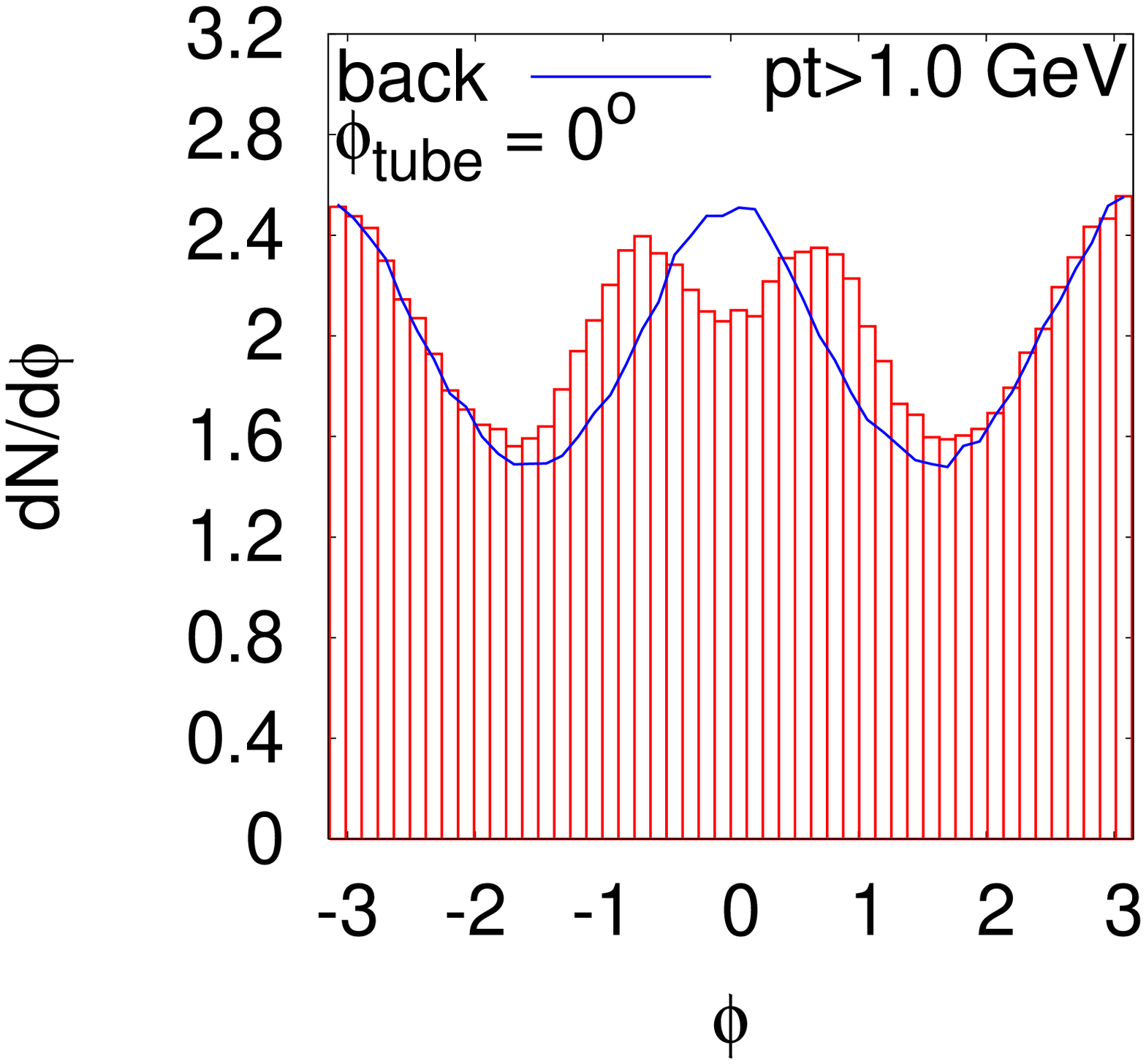}
\hspace{-.6cm}
\includegraphics[width=4.5cm]{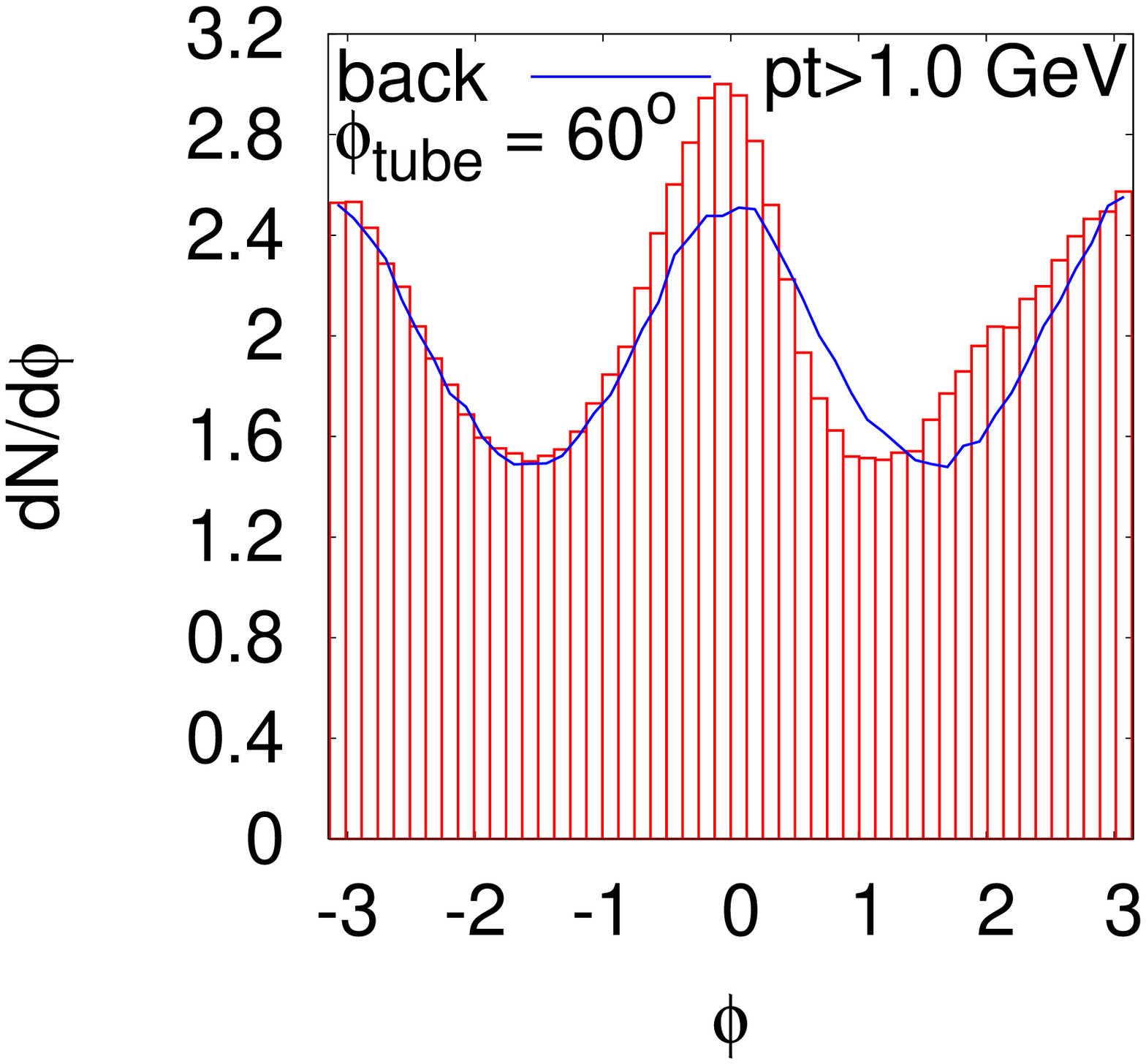}
\caption{\label{noncentr} Tube model for a non-central  
 collision (20-30\% centrality). From left to right: 
 Energy density distribution, single-particle angular 
 distribution for tube in the reaction plane, idem but tube  
 at $60^o$.}
\end{figure}

\begin{figure}[h] 
\begin{minipage}[h]{60mm} 
\hspace{-.3cm} 
\includegraphics[width=6.cm]{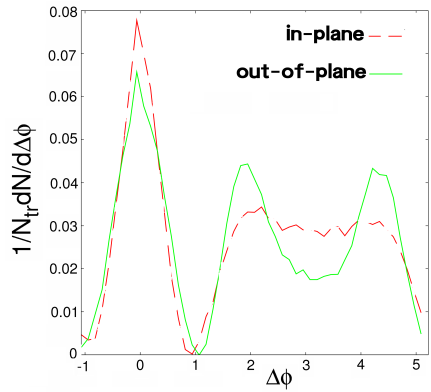}
\vspace*{-.3cm} 
\caption{\label{nccor} Two-particle correlation for 
 different trigger (not tube) angle.} 
\end{minipage} 
\hspace{1.cm}
\begin{minipage}[t]{63mm} 
\vspace*{-3.5cm} 
\hspace{.5cm} With these information, we can discuss what happens in a more complex event such as a NeXus event. In 
such an event, only the outer tubes need to be considered. 
The shape of the two-particle correlations for a single tube  
(in particular the peak spacing) is relatively independent
of its features so the various tubes will contribute with 
rather similar two-peaks emission pattern at various angles 
in the single-particle angular distribution. For this single 
event, the two-particle correlation has a well-defined main 
structure similar to that of a single tube (Fig. \ref{1dist}) 
surrounded by several other peaks and depressions due to 
trigger and associated particles coming from different tubes. 
These additional\hfilneg\ 
\end{minipage} 
\vspace{-.4cm} 
\end{figure}

The results above were obtained for central collisions. 
We have also generalized this one tube model to non-central 
collisions. In Fig.~\ref{noncentr}, the single-particle 
angular distribution is shown with and without tube. When 
the tube is present, the single particle angular 
distribution may also have two peaks with separation $\Delta\phi\sim 2$ or this structure may be more 
hidden by the elliptic flow depending on the tube position. 
After subtracting the elliptic flow and averaging over the 
tube angular position, the two-particle correlation has a 
near-side peak and an away-side substructure that is indeed 
double-peaked for more  central collisions and single-peak 
for more peripheral 
ones. We have also studied the dependence 
on the trigger angle with respect to the reaction plane 
and found the correct behavior (single peak for in-plane 
trigger and double peak for out-of-plane one). This result 
is shown in Fig. \ref{nccor}.

\noindent 
peaks and depressions have positions depending on the angle 
of the tubes between them. When averaging over many randomly 
fluctuating events these interference terms disappear and 
only the main one-tube like structure is left.

\section{Conclusion} 

To summarize,  a unified picture for the  structures observed 
in two particle correlations at low to moderate transverse 
momentum was presented. It is based on the presence of 
longitudinal high energy density tubes in the initial 
conditions. These tubes are leftover from the initial 
particle interactions. During the hydrodynamical evolution 
of the fluid, pressed by the strong expansion of the tubes 
located close to the border, matter (coming mainly from the 
background) is deflected into two symmetrical directions, 
leading to  two-particle correlations with a near-side and a 
double hump away-side ridges. 

\begin{figure}[h] 
\vspace*{-.2cm} 
\begin{minipage}[h]{85mm}
\hspace{-.5cm} 
\includegraphics[width=4.cm]{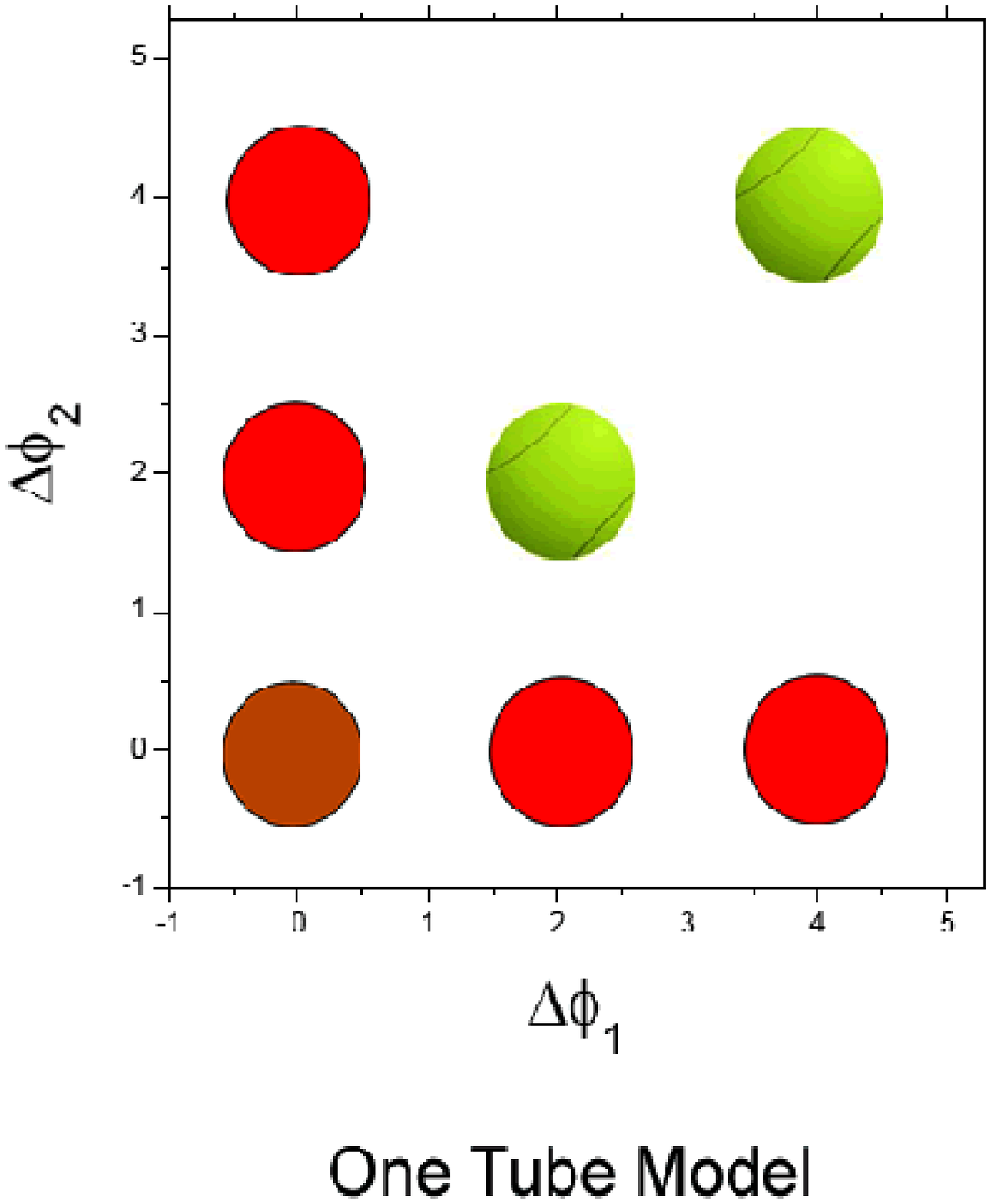} 
\hspace{-.7cm} 
\includegraphics[width=5.5cm]{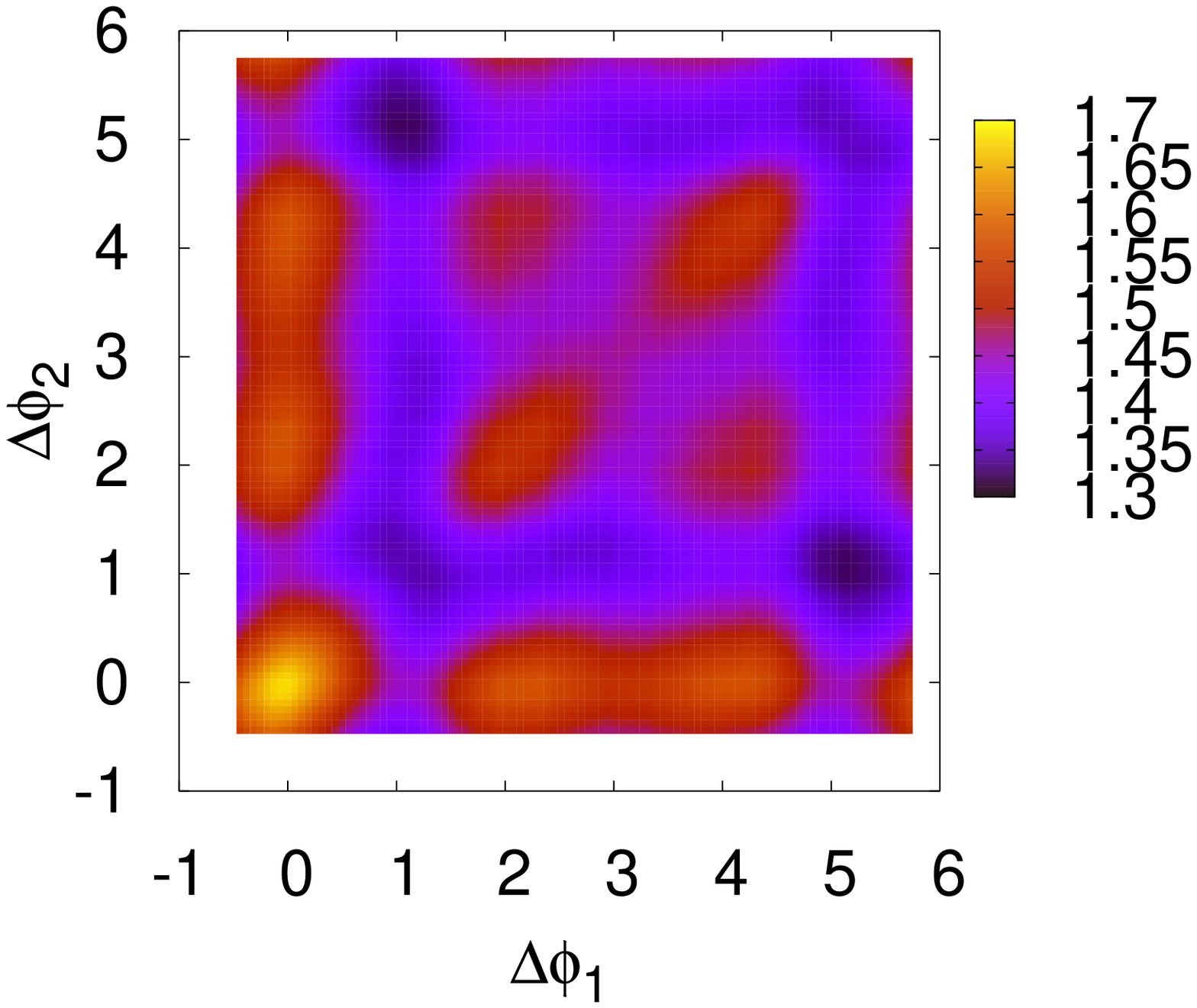}
\caption{\label{3pcorr} Three-particle correlation in $\Delta \phi_1$-$\Delta \phi_2$:
 naive expectation for the one-tube model and computed raw correlation.} 
\end{minipage} 
\hspace{.8cm} 
\begin{minipage}[t]{41mm} 
\vspace*{-3.3cm}  
\hspace{.4cm}
Another possibility would be 2+1 correlations, where the 
trigger and first associated particle are chosen in different 
peaks of the single particle angular distribution 
($\Delta \phi_1 \sim 2$) and the second associated particle 
position varies. When this correlation is plotted as 
function of $\Delta \phi_2$ and $\Delta \eta_2$ 
(pseudorapidity difference with respect to the trigger), 
there should appear two main structures elongated in $\Delta \eta_2$ corresponding to the two possibilities for\hfilneg\ 
\end{minipage} 
\vspace{-.4cm} 
\end{figure}

Various models have been suggested to explain the structures 
in the two particle correlations and even those based on 
hydrodynamics have strong differences among them: for each 
tube, there may be no emission \cite{shuryakhole}, emission 
in one direction \cite{lml1,lml2,sg} or two directions as 
advocated here. Three-particle correlations have been 
suggested as a possible test to distinguish between various 
scenarios. For example, three-particle correlations can be 
plotted as function of the angles $\Delta\phi_1$ between 
the first associated particle  and the trigger and 
$\Delta\phi_2$ between the second associated particle and 
the trigger. The Mach cone model \cite{renk} or the AMPT 
model \cite{ma} would lead to off diagonal peaks in this 
plot while certain models might not. Experimentally, such off 
diagonal peaks do appear in three-particle correlations 
\cite{star3} (see also the approach in \cite{phenix3}).  
Naively, in our case, one might expect no off diagonal 
peaks because there are only two peaks in the single 
particle angular distribution (Fig. \ref{3pcorr} left). 
However if one carefully consider the contributions coming 
from different 3-dimensional integration domains (in 
$\phi_t\,$, $\phi_1$ and $\phi_2$), 
there appear also two off-diagonal local maxima in the 
three-particle correlation. This can be seen in the computed 
result in Fig. \ref{3pcorr} right. So this type of 
correlation does not look promising to test our model.  

\noindent 
the second 
associated particle: it comes from the same peak as the 
trigger or same peak as the first associated particle. Though 
it is not ruled out, it seems difficult that jet based models 
lead to this. 

To conclude: usually, the initial conditions in the 
hydrodynamic description of relativistic nuclear collisions 
are assumed to be smooth. It seems however that each time 
more data require a knowledge of the event-by-event 
fluctuating initial conditions rather than smooth initial 
conditions (for example fluctuations in elliptic flow 
\cite{spheriofluct} or perhaps the very behavior of elliptic 
flow as function of pseudorapity \cite{spheriov2}). Now 
two-particle correlations might also offer us a chance to 
get a glimpse of the initial conditions and learn about the 
strong interaction. 

We acknowledge funding from 
 CNPq and FAPESP.





\bibliographystyle{model1a-num-names}
\bibliography{ridge}

\providecommand{\noopsort}[1]{}\providecommand{\singleletter}[1]{#1}%
\begin{thebibliography}{36}
\expandafter\ifx\csname natexlab\endcsname\relax\def\natexlab#1{#1}\fi
\providecommand{\bibinfo}[2]{#2}
\ifx\xfnm\relax \def\xfnm[#1]{\unskip,\space#1}\fi
\bibitem[{{J. Putschke for the STAR
  collaboration}(2007{\natexlab{a}})}]{ridge1}
\bibinfo{author}{{J. Putschke for the STAR collaboration}},
  \bibinfo{journal}{Nucl.\ Phys. \ A} \bibinfo{volume}{783}
  (\bibinfo{year}{2007}{\natexlab{a}}) \bibinfo{pages}{507}.
\bibitem[{{J. Putschke for the STAR
  collaboration}(2007{\natexlab{b}})}]{ridge1s}
\bibinfo{author}{{J. Putschke for the STAR collaboration}},
  \bibinfo{journal}{J.\ Phys. \ G} \bibinfo{volume}{34}
  (\bibinfo{year}{2007}{\natexlab{b}}) \bibinfo{pages}{S679}.
\bibitem[{{M.P. McCumber for the PHENIX Collaboration}(2007)}]{ridge2}
\bibinfo{author}{{M.P. McCumber for the PHENIX Collaboration}},
  \bibinfo{journal}{J.\ Phys. \ G} \bibinfo{volume}{35} (\bibinfo{year}{2007})
  \bibinfo{pages}{104081}.
\bibitem[{{M.J. Horner for the STAR Collaboration}(2007)}]{dhump1}
\bibinfo{author}{{M.J. Horner for the STAR Collaboration}},
  \bibinfo{journal}{J.\ Phys. \ G} \bibinfo{volume}{34} (\bibinfo{year}{2007})
  \bibinfo{pages}{S995}.
\bibitem[{{E. Wenger for the PHOBOS Collaboration}(2008)}]{phobos}
\bibinfo{author}{{E. Wenger for the PHOBOS Collaboration}},
  \bibinfo{journal}{J.\ Phys. \ G} \bibinfo{volume}{35} (\bibinfo{year}{2008})
  \bibinfo{pages}{104080}.
\bibitem[{{B. Alver et al. (PHOBOS Collaboration)}(2010)}]{phoboss}
\bibinfo{author}{{B. Alver et al. (PHOBOS Collaboration)}},
  \bibinfo{journal}{Phys. \ Rev. \ Lett.} \bibinfo{volume}{104}
  (\bibinfo{year}{2010}) \bibinfo{pages}{062301}.
\bibitem[{{M. Daugherity for the STAR Collaboration}(2008)}]{ridge3}
\bibinfo{author}{{M. Daugherity for the STAR Collaboration}},
  \bibinfo{journal}{J. \ Phys. \ G} \bibinfo{volume}{35} (\bibinfo{year}{2008})
  \bibinfo{pages}{104090}.
\bibitem[{Voloshin(2006)}]{voloshin}
\bibinfo{author}{S.~Voloshin}, \bibinfo{journal}{Phys. \ Lett.\ B}
  \bibinfo{volume}{632} (\bibinfo{year}{2006}) \bibinfo{pages}{490}.
\bibitem[{Shuryak(2007)}]{shuryak}
\bibinfo{author}{E.~Shuryak}, \bibinfo{journal}{Phys. \ Rev. \ C}
  \bibinfo{volume}{76} (\bibinfo{year}{2007}) \bibinfo{pages}{047901}.
\bibitem[{{A. Dumitru et al.}(2008)}]{lml1}
\bibinfo{author}{{A. Dumitru et al.}}, \bibinfo{journal}{Nucl. \ Phys. \ A}
  \bibinfo{volume}{810} (\bibinfo{year}{2008}) \bibinfo{pages}{91}.
\bibitem[{{S. Gavin et al.}(2009)}]{lml2}
\bibinfo{author}{{S. Gavin et al.}}, \bibinfo{journal}{Phys. \ Rev. \ C}
  \bibinfo{volume}{79} (\bibinfo{year}{2009}) \bibinfo{pages}{051902}.
\bibitem[{{G. Moschelli and S. Gavin}(????)}]{sg}
\bibinfo{author}{{G. Moschelli and S. Gavin}},
  \bibinfo{journal}{arXiv:0910.3590v2} .
\bibitem[{{T. Hirano and K. Tsuda}(2002)}]{Hirano}
\bibinfo{author}{{T. Hirano and K. Tsuda}}, \bibinfo{journal}{Phys. \ Rev. \ C}
  \bibinfo{volume}{66} (\bibinfo{year}{2002}) \bibinfo{pages}{054905}.
\bibitem[{{C. Nonaka and S.A. Bass}(2007)}]{Nonaka}
\bibinfo{author}{{C. Nonaka and S.A. Bass}}, \bibinfo{journal}{Phys. \ Rev. \
  C} \bibinfo{volume}{75} (\bibinfo{year}{2007}) \bibinfo{pages}{014902}.
\bibitem[{{C.E. Aguiar, Y. Hama, T. Kodama and T.
  Osada}(2001{\natexlab{a}})}]{spherioref1}
\bibinfo{author}{{C.E. Aguiar, Y. Hama, T. Kodama and T. Osada}},
  \bibinfo{journal}{J. \ Phys. G} \bibinfo{volume}{27}
  (\bibinfo{year}{2001}{\natexlab{a}}) \bibinfo{pages}{75}.
\bibitem[{{C.E. Aguiar, Y. Hama, T. Kodama and T.
  Osada}(2001{\natexlab{b}})}]{spherioref2}
\bibinfo{author}{{C.E. Aguiar, Y. Hama, T. Kodama and T. Osada}},
  \bibinfo{journal}{J. \ Phys. G} \bibinfo{volume}{27}
  (\bibinfo{year}{2001}{\natexlab{b}}) \bibinfo{pages}{551}.
\bibitem[{{C.E. Aguiar, Y. Hama, T. Kodama and T. Osada}(2002)}]{spherioref3}
\bibinfo{author}{{C.E. Aguiar, Y. Hama, T. Kodama and T. Osada}},
  \bibinfo{journal}{Nucl. \ Phys. \ A} \bibinfo{volume}{698}
  (\bibinfo{year}{2002}) \bibinfo{pages}{639c}.
\bibitem[{{O. Socolowski Jr., F. Grassi, Y. Hama and T. Kodama}(2004)}]{sph1}
\bibinfo{author}{{O. Socolowski Jr., F. Grassi, Y. Hama and T. Kodama}},
  \bibinfo{journal}{Phys. \ Rev. \ Lett.} \bibinfo{volume}{93}
  (\bibinfo{year}{2004}) \bibinfo{pages}{18230}.
\bibitem[{{ R. Andrade, F. Grassi, Y. Hama, O. Socolowski Jr. and T.
  Kodama}(2006)}]{sph2}
\bibinfo{author}{{ R. Andrade, F. Grassi, Y. Hama, O. Socolowski Jr. and T.
  Kodama}}, \bibinfo{journal}{Phys. \ Rev. \ Lett.} \bibinfo{volume}{97}
  (\bibinfo{year}{2006}) \bibinfo{pages}{202302}.
\bibitem[{{ R.P.G. Andrade, F. Grassi, Y. Hama, T. Kodama and W.-L.
  Qian}(2008)}]{sph3}
\bibinfo{author}{{ R.P.G. Andrade, F. Grassi, Y. Hama, T. Kodama and W.-L.
  Qian}}, \bibinfo{journal}{Phys. \ Rev. \ Lett.} \bibinfo{volume}{101}
  (\bibinfo{year}{2008}) \bibinfo{pages}{112301}.
\bibitem[{{J. Takahashi, B.M. Tavares, W.-L. Qian, R. Andrade, F. Grassi, Y.
  Hama, T. Kodama, and N. Xu}(2009)}]{jun}
\bibinfo{author}{{J. Takahashi, B.M. Tavares, W.-L. Qian, R. Andrade, F.
  Grassi, Y. Hama, T. Kodama, and N. Xu}}, \bibinfo{journal}{Phys. \ Rev. \
  Lett.} \bibinfo{volume}{103} (\bibinfo{year}{2009}) \bibinfo{pages}{242301}.
\bibitem[{{H. Petersen, J. Steinheimer, G. Burau, M. Bleicher, H.
  St\"ocker}(2008)}]{urqmdhydro}
\bibinfo{author}{{H. Petersen, J. Steinheimer, G. Burau, M. Bleicher, H.
  St\"ocker}}, \bibinfo{journal}{Phys. \ Rev.\ C} \bibinfo{volume}{78}
  (\bibinfo{year}{2008}) \bibinfo{pages}{044901}.
\bibitem[{{K. Werner, Iu. Karpenko, T. Pierog, M. Bleicher and K.
  Mikhailov}(????)}]{klaus}
\bibinfo{author}{{K. Werner, Iu. Karpenko, T. Pierog, M. Bleicher and K.
  Mikhailov}}, \bibinfo{journal}{arXiv:1004.0805} .
\bibitem[{{H.J. Drescher, F.M. Liu, S. Ostapchenko, T. Pierog and K.
  Werner}(2002)}]{nexus}
\bibinfo{author}{{H.J. Drescher, F.M. Liu, S. Ostapchenko, T. Pierog and K.
  Werner}}, \bibinfo{journal}{Phys.\ Rev.\ C} \bibinfo{volume}{65}
  (\bibinfo{year}{2002}) \bibinfo{pages}{054902}.
\bibitem[{{Y. Hama, R.P.G. Andrade, F. Grassi and W.-L. Qian}(2010)}]{ismd09}
\bibinfo{author}{{Y. Hama, R.P.G. Andrade, F. Grassi and W.-L. Qian}},
  \bibinfo{journal}{Nonlin. \ Phenom.\ Complex \ Sys.} \bibinfo{volume}{12}
  (\bibinfo{year}{2010}) \bibinfo{pages}{446}.
\bibitem[{{R.P.G. Andrade, F. Grassi, Y. Hama and W.-L. Qian}(????)}]{sqm09}
\bibinfo{author}{{R.P.G. Andrade, F. Grassi, Y. Hama and W.-L. Qian}},
  \bibinfo{journal}{{arXiv:0912.0703}} .
\bibitem[{Werner(????)}]{KWprivate}
\bibinfo{author}{K.~Werner}, \bibinfo{journal}{private communication} .
\bibitem[{{A. Adare et al. (PHENIX Collaboration)}(2008)}]{dhump3}
\bibinfo{author}{{A. Adare et al. (PHENIX Collaboration)}},
  \bibinfo{journal}{Phys. \ Rev. \ C} \bibinfo{volume}{78}
  (\bibinfo{year}{2008}) \bibinfo{pages}{014901}.
\bibitem[{{ A. Feng (for the STAR collaboration)}(2008)}]{inout}
\bibinfo{author}{{ A. Feng (for the STAR collaboration)}}, \bibinfo{journal}{J.
  \ Phys.\ G} \bibinfo{volume}{35} (\bibinfo{year}{2008})
  \bibinfo{pages}{104082}.
\bibitem[{{E. Shuryak}(2009)}]{shuryakhole}
\bibinfo{author}{{E. Shuryak}}, \bibinfo{journal}{Phys. \ Rev. \ C}
  \bibinfo{volume}{80} (\bibinfo{year}{2009}) \bibinfo{pages}{054908}.
\bibitem[{{T. Renk and J. Ruppert}(2007)}]{renk}
\bibinfo{author}{{T. Renk and J. Ruppert}}, \bibinfo{journal}{Phys. \ Rev. \ C}
  \bibinfo{volume}{76} (\bibinfo{year}{2007}) \bibinfo{pages}{014908}.
\bibitem[{{G.L. Ma et al}(2007)}]{ma}
\bibinfo{author}{{G.L. Ma et al}}, \bibinfo{journal}{Phys. \ Lett. \ B}
  \bibinfo{volume}{647} (\bibinfo{year}{2007}) \bibinfo{pages}{122}.
\bibitem[{{B.I. Abelev et al. (STAR collaboration)}(2009)}]{star3}
\bibinfo{author}{{B.I. Abelev et al. (STAR collaboration)}},
  \bibinfo{journal}{Phys. \ Rev. \ Lett.} \bibinfo{volume}{102}
  (\bibinfo{year}{2009}) \bibinfo{pages}{052302}.
\bibitem[{{N.N. Ajitanand for the PHENIX collaboration}(2007)}]{phenix3}
\bibinfo{author}{{N.N. Ajitanand for the PHENIX collaboration}},
  \bibinfo{journal}{Nucl. \ Phys. \ A} \bibinfo{volume}{783}
  (\bibinfo{year}{2007}) \bibinfo{pages}{519}.
\bibitem[{{Y. Hama, R.P.G. Andrade, F. Grassi, W.-L. Qian, T. Osada, C.E.
  Aguiar and T. Kodama}(2008)}]{spheriofluct}
\bibinfo{author}{{Y. Hama, R.P.G. Andrade, F. Grassi, W.-L. Qian, T. Osada,
  C.E. Aguiar and T. Kodama}}, \bibinfo{journal}{Phys. \ Atom. \ Nucl.}
  \bibinfo{volume}{71} (\bibinfo{year}{2008}) \bibinfo{pages}{1558}.
\bibitem[{{R.P.G. Andrade, F. Grassi, Y. Hama, T. Kodama and W.-L.
  Qian}(2008)}]{spheriov2}
\bibinfo{author}{{R.P.G. Andrade, F. Grassi, Y. Hama, T. Kodama and W.-L.
  Qian}}, \bibinfo{journal}{Phys. \ Rev. \ Lett.} \bibinfo{volume}{101}
  (\bibinfo{year}{2008}) \bibinfo{pages}{112301}.

\end{thebibliography}







\end{document}